\documentstyle[12pt]{article}

\newcommand{\be}{\begin{equation}}
\newcommand{\ee}{\end{equation}}
\newcommand{\ba}{\begin{eqnarray}}
\newcommand{\ea}{\end{eqnarray}}
\newcommand{\bas}{\begin{eqnarray*}}
\newcommand{\eas}{\end{eqnarray*}}

\newcommand{\eq}[1]{{(\ref{#1})}}

\newcommand{\nn}{\nonumber}
\newcommand{\g}{\gamma}
\newcommand{\Lam}{\Lambda}

\newcommand{\dl}{\delta}

\newcommand{\oL}{\frac{1}{L}}

\begin{document}
\topmargin 0pt
\oddsidemargin 5mm
\headheight 0pt
\topskip 0mm
 
\hfill UT-Komaba 97-4

\hfill YITP-97-9

\hfill February 1997

\begin{center}
 
\vspace{36pt}
{\large \bf  Boundary bound states for \\
the open Hubbard chain with boundary fields}
 
\end{center}

\vspace{36pt}
 
\begin{center}

\vspace{5pt}
 Osamu Tsuchiya$^1$ and Takashi Yamamoto$^2$\\
\vspace{3pt}
{\sl $^1$Department of Pure and Applied Sciences, 
University of Tokyo, \\
Komaba, Meguro-ku, Tokyo 153, Japan}

\vspace{3pt}

\vspace{3pt}
{\sl  $^2$Yukawa Institute for Theoretical Physics, 
Kyoto University, \\
Kyoto 606, Japan}
 
\end{center}

\vspace{36pt}
 
 
\begin{center}
{\bf Abstract}
\end{center}
 
\vspace{12pt}
 
\noindent
The boundary effects in the open Hubbard chain with boundary fields
are studied.
The boundary string solutions of the Bethe ansatz equations 
that give rise to a wave functions localized at the boundary
and exponentially decreasing away from the boundary
are provided.
In particular, it is shown that
the correct ground state of the model at half-filling 
contains the boundary strings.

\vspace{24pt}
 
\vfill
 
\newpage


In view of the experimental possibilities \cite{experm},
interest in the theoretical study of 
transport properties for one-dimensional electron systems
has been considerably enhanced during recent years
(see ref. \cite{rev-trans} and references therein).
Thus, theoretical investigations of Tomonaga-Luttinger liquids 
(see \cite{TLL} and references therein) 
in the presence of boundaries and potential scatterers
are of crucial importance.
A conspicuous example of Tomonaga-Luttinger liquids
with boundaries is provided by 
the one-dimensional Hubbard model with open boundaries 
and boundary fields (open Hubbard chain with boundary fields)
\cite{S85,AS,SW,DY}.
The Hamiltonian of this model is given by
$H
=H_{\mbox{{\scriptsize bulk}}}^{(\mbox{{\scriptsize open}})}
+H_{\mbox{{\scriptsize boundary}}}$,
where 
\ba
\label{bulk-hamil}
H_{\mbox{{\scriptsize bulk}}}^{(\mbox{{\scriptsize open}})} 
&=& 
-\sum_{j=1}^{L-1}\sum_{\sigma=\uparrow,\downarrow}
(\psi_{j \sigma}^{\dagger}\psi_{j+1 \sigma} 
+\psi_{j+1 \sigma}^{\dagger}\psi_{j \sigma})
+ 
U\sum_{j=1}^{L}
n_{j \uparrow}n_{j\downarrow}, \\ 
\label{boundary-hamil}
H_{\mbox{{\scriptsize boundary}}}
&=&
-h_{1}(n_{1\uparrow}+ n_{1\downarrow})
-h_{L}(n_{L\uparrow}+ n_{L\downarrow}). 
\ea
Here $U$ is the coupling constant, $h_{l}$ is the boundary
field at site $l\in\{1,L\}$,
$\psi_{j \sigma}$ (resp. $\psi_{j\sigma}^{\dagger}$)
denotes the annihilation (resp. the creation) operator 
of an electron with spin $\sigma \in \{\uparrow, \downarrow \}$
at site $j\in\{1,2,\cdots, L\}$,
and $n_{j \sigma} =\psi_{j \sigma}^{\dagger} \psi_{j \sigma}$
is the number operator.
This model has been solved exactly by Bethe ansatz \cite{S85,AS,SW,DY},
and its long-distance properties have been analyzed 
by using boundary conformal field theory \cite{AS,SW,DY}.
Also, the boundary $S$ matrices of the model have been obtained \cite{Ts}.

In this letter, focusing on the complex solutions of 
the Bethe ansatz equations, 
we proceed to study the boundary effects in 
the open Hubbard chain with boundary fields.
Originally, such kind of work has been done in the case
of open $XXZ$ chain with boundary magnetic fields \cite{SS95,KS96}.
In refs. \cite{SS95,KS96},
the authors found {\it boundary string solutions}
to the Bethe ansatz equations. 
The boundary strings represent {\it boundary bound states}
which correspond to the wave functions localized at
the boundaries.
In particular, it was shown that, 
in the gapfull regime with appropriate 
conditions \cite{JK} on the boundary magnetic fields,
the ground state contains the boundary 1-string.

We will show that, for the open Hubbard chain with boundary fields of 
type (\ref{boundary-hamil}),
there exist boundary bound states of the electrons 
corresponding to each boundary fields.
If the boundary fields is sufficiently large
($h_{l}>1$ for $l=1,L$)
then the rapidity configuration in the ground state of this model 
contains boundary strings.

Let us recall the Bethe ansatz solutions of 
the open Hubbard chain \cite{S85,AS,SW,DY}.
The energy is represented as
\ba
E & = & -2 \sum_{j=1}^{N} \cos k_{j}, \label{energy}
\ea
where the charge rapidities $k_{j}$ are the solutions
of the Bethe ansatz equations

\ba
\label{bethe-eq1} 
e^{i2k_{j}(L+1)} 
\beta(k_{j},h_{1}) \beta(k_{j},h_{L})=  
\prod_{\delta =1}^{M}
e\left( 2(\sin k_{j} -\Lam_{\delta})/c \right)
e\left( 2(\sin k_{j} +\Lam_{\delta})/c \right), &&\\
\label{bethe-eq2} 
\prod_{{\scriptstyle \dl=1} \atop {\scriptstyle \dl (\neq \g)}}^{M}
e\left( (\Lam_{\gamma} -\Lam_{\delta} )/c \right) 
e\left( (\Lam_{\gamma} +\Lam_{\delta} )/c \right) 
=
\prod_{j=1}^{N} 
e\left( 2(\Lam_{\gamma} -\sin k_{j} )/c \right) 
e\left( 2(\Lam_{\gamma} +\sin k_{j} )/c \right). 
\ea 
Here $N$ and $M$ respectively denote the number of electrons and electrons with 
down spin, and 
\ba
&&
c=U/2,  \\
&&
\beta(x,h)=\frac{1-he^{-ix}}{1-he^{ix}},\\
&&
e(x)= \frac{x+i}{x-i}. 
\ea
Note that, in this model,
the solutions of the Bethe ansatz equations
are restricted as $\mbox{Re}(k_j), \mbox{Re}(\Lam_\gamma)\geq 0$
and $k_j, \Lam_\gamma\ne 0$.

Putting $k_{-j} =-k_{j}$,$\Lam_{-\gamma}=-\Lam_{\gamma}$
and $k_{0}=0,\Lam_{0}=0$, we can rewrite the above Bethe ansatz equations
into the more tractable forms,
\ba
\label{doubl-bethe-eq1} 
e^{i2k_{j}(L+1)} 
\beta(k_{j},h_{1}) \beta(k_{j},h_{L})
=  
e\left( 2\sin k_{j}/c \right)^{-1}
\prod_{\delta =-M}^{M}
e\left( 2(\sin k_{j} -\Lam_{\delta})/c \right), &&\\
\label{doubl-bethe-eq2} 
e\left( \Lam_{\gamma}/c \right)^{-1}
\prod_{{\scriptstyle \dl=-M} \atop {\scriptstyle \dl (\neq \g)}}^{M}
e\left( (\Lam_{\gamma} -\Lam_{\delta} )/c \right) 
=
\prod_{j=-N}^{N} 
e\left( 2(\Lam_{\gamma} -\sin k_{j} )/c \right).
\ea 
Also, the energy is represented as
\ba
E & = & - \sum_{j=-N}^{N} \cos k_{j}+1. \label{doubl-energy}
\ea

For the  problems with periodic boundary conditions,
we usually adopt the string hypothesis \cite{T72}.
The string hypothesis states  that, 
in the thermodynamic limit,
the set of solutions 
$\{k_{j},\Lam_{\gamma}\}$ of the Bethe ansatz 
equations can be split into three kinds of solutions,
which are
real $k_{j}$'s, 
combination of $n$  $\Lam_{\gamma}$'s ($\Lam$-strings of length $n$)
and combination of $n$ $\Lam_{\gamma}$'s and 
$2n$ $k_{j}$'s ($k$-$\Lam$-strings of length $n$)
\cite{T72}. 
(We shall call these the {\it bulk} string solutions.)
The $\Lam$-strings and the $k$-$\Lam$-strings can be interpreted
as some kind of bound states \cite{T72,EKS92}.

For the problems with open boundary conditions,
since the equations (\ref{doubl-bethe-eq1}) and (\ref{doubl-bethe-eq2})
are very similar to the one 
for the model with periodic boundary conditions,
it is reasonable to conjecture that the bulk strings
make up the (sub)set of the 
solutions of the Bethe ansatz equations.

\noindent
{\it Boundary string}

For the model with open boundary conditions, however, 
there may exist the complex solutions
which are the different kind to the bulk strings.
Note that, for the model with open boundary conditions,
the energy (\ref{energy}) is the integrals of motion but the 
total momentum is not. 
Thus $\sum_{j=1}^{N} k_{j}$ does not have to be real,
in contrast to the case of the periodic boundary conditions. 
The rapidity configurations which contain pure imaginary roots
(or roots whose real parts are $\pi$)
and do not contain the complex conjugate of these pure imaginary roots
are admissible.
(We shall call this kind of solutions which 
correspond to the boundary fields \lq{\it boundary string solutions}\rq.
\cite{SS95,KS96})

\noindent
{\it One particle systems}

As a warm-up exercise, we first consider the one electron
system ($N=1$). 
The wave function of the Bethe ansatz states is given by
\ba
\Phi_{\sigma}(n) & = A_{\sigma} (k) e^{ikn} - B_{\sigma}(k) e^{-ikn}
\ea
where $\sigma = \uparrow $ or $\downarrow$ is the spin 
of an electron.
The relation between $A_{\sigma}(k)$ and $B_{\sigma}(k)$ 
is obtained by investigating 
the boundary scattering at 
left or right boundaries.
The rapidity $k$ is determined
by the consistency condition for the left and right scatterings
(Bethe ansatz equation). 
The boundary scattering at the boundary $n=1$ gives
\ba
&&
B_{\sigma} (k)  =  
\frac{1-h_{1}e^{ik}}{1-h_{1}e^{-ik}} A_{\sigma} (k)  
\label{bdryscattering-left}
\ea
and at $n=L$ gives
\ba
&&
B_{\sigma}(k) = \frac{1-h_{L}e^{-ik}}{1-h_{L}e^{ik}}
               e^{ik2(L+1)} A_{\sigma}(k).
\label{bdryscattering-right}
\ea
The consistency condition for both left and right scatterings
gives
\ba
&&
\frac{1-h_{1} e^{ik} }{1-h_{1} e^{-ik}} 
\frac{1-h_{L} e^{ik} }{1-h_{L} e^{-ik}}   =  
e^{2ik(L+1)}. \label{bethe-1}
\ea
There are two type of solutions for  (\ref{bethe-1}):
one is the real solution which correspond to  the
free particle interacting at the boundary,
the other is the pure imaginary solution 
which can be interpreted as the bound state at the boundary.
When $k$ is pure imaginary, the right hand side of 
(\ref{bethe-1}) decreases exponentially fast  in the thermodynamic
limit (we deal with only the case $\mbox{Im}(k) >0$).
Therefore, if eq. (\ref{bethe-1}) has pure imaginary solution,
the rapidity $k$ must take values of which the left hand side of 
(\ref{bethe-1}) vanishes. 
Let $k_{(1)} = i\chi_{(1)}, k_{(L)} = i\chi_{(L)}$ be
the pure imaginary solutions such as 
\ba
&&
1-e^{-\chi_{(1)}} h_{1} \sim e^{-2 \chi_{(1)} (L+1)}, \nn \\
&& 1-e^{-\chi_{(L)}} h_{L} \sim e^{-2 \chi_{(L)} (L+1)}. 
\ea
Note that the above equations have 
solution with $\chi_{(l)} >0$
only when $h_l>1$ $(l=1,L)$.
The solution $k_{(1)}$ ($k_{(L)}$) 
corresponds to the
wavefunction localized at the site $1$: 
$\Phi_{\sigma} \sim e^{-n\chi_{(1)}}$
(resp. site $L$:
$\Phi_{\sigma} \sim e^{-(L-n)\chi_{(L)}}$ )



\noindent
{\it Ground state}

Now we investigate the ground state 
of the open Hubbard chain with boundary fields
at the half filling%
\footnote{In this letter, the term half-filling
         means that the real charge rapidities
         $k_{j}$ fill the Fermi-sea  
        between $-\pi$ and $\pi$. }.
For the model with periodic boundary conditions, 
in the ground state,
the rapidities $\{k_{j},\Lam_{\gamma} \}$
are all real and fill the Fermi seas.
However, as examined above, we must consider 
the possibility of the existence of the boundary strings.
Our task is, then, to compare the energy of the configuration 
with boundary strings to 
that of the configuration without the boundary strings.
(In the remainder of this letter, we treat
only the case that the real roots $\{k_{j},\Lam_{\gamma} \}$
fill the Fermi-seas.) 
To calculate the effect of the existence of the boundary 
strings to the energy, 
we must investigate, in addition to the
energy of the boundary strings, 
the shift of the densities of the real roots due to
the existence of the boundary strings.

We shall introduce the densities of roots and holes.
The number of allowed solutions for the Bethe ansatz equations 
(\ref{bethe-eq1}) and (\ref{bethe-eq2}) 
in the intervals ($k$, $k+dk$) and ($\Lam$, $\Lam+d\Lam$)
are expressed as 
$L[\rho(k)+\rho^{h}(k)]dk$ and 
$L[\sigma(\Lam)+\sigma^{h}(\Lam)]d\Lam$.
Here $\rho(k)$ and $\sigma(\Lam )$ 
are the densities of roots (filled solutions),
and $\rho^{h}(k)$ and $\sigma^{h}(\Lam)$ 
are the densities of holes (unfilled solutions).

If there exist two boundary strings $\chi_{(1)}=\ln h_{1}$,
$ \chi_{(L)}=\ln h_{L}$ 
for each boundary fields
(and there do not exist holes 
in the real roots), 
the densities $\rho(k)$, $\sigma(\Lam)$ satisfy 
the following integral equations
\ba
\rho(k) 
&=& 
\frac{1}{\pi}
+2\cos k \int_{-B}^{B}d\Lam\sigma(\Lam)K(2(\sin k -\Lam)) \nn \\
& &+\frac{1}{L}
\left[
\frac{1}{2\pi}(\xi(k,h_1)+\xi(k,h_L))
-2\cos k K(2\sin k) 
\right],\\
\sigma (\Lam ) 
& = & 
2 \int_{-Q}^{Q} dk \rho (k) K(2 \Lam -2 \sin k) 
- \int_{-B}^{B} d \Lam'  \sigma(\Lam') K(\Lam -\Lam ') \nn \\
& &
+\frac{1}{L}
\left[ \frac{2c}{c-2\sinh \chi_{(1)}} 
K(\frac{2c\Lam}{c-2\sinh \chi_{(1)}})
+
\frac{2c}{c+2\sinh \chi_{(1)} }
K(\frac{2c\Lam}{c+2\sinh \chi_{(1)}}) \right] \nn \\
&&
+\frac{1}{L}
\left[ \frac{2c}{c-2\sinh \chi_{(L)}} 
K(\frac{2c\Lam}{c-2\sinh \chi_{(L)}})
+
\frac{2c}{c+2\sinh \chi_{(L)} }
K(\frac{2c\Lam}{c+2\sinh \chi_{(L)}}) \right] \nn \\
&&
-\frac{1}{L} K (\Lam),
\ea
where $\xi(k,h)=(h^{2}-1)/(1-2h \cos k +h^{2})$,
$K(x)= c/[\pi(x^2 +c^2 )]$. 
Here $Q$ and $B$ are charge and spin pseudo Fermi-momentum, respectively.
We assume that, to examine the densities of order $O(1/L)$ 
in the half filled case,
we can take $B=\pi$ and $Q=\infty$.

If we separate the densities into the part of  order $O(L^{0})$ and 
$O(1/L)$ as
\ba
\rho (k) & = & \rho_{0} (k) 
+ \frac{1}{L} \rho_{1} (k) +O(\frac{1}{L^{2}})\nn
\\
\sigma (\Lam ) & = & \sigma_{0} (\Lam) 
+\frac{1}{L} \sigma_{1} (\Lam)+O(\frac{1}{L^{2}}),
\ea
then $\rho_{0}(k)$ and $\sigma_{0}(\Lam)$ are identical to 
the densities of $k$'s 
and $\Lam$'s, respectively, for 
the ground state with periodic boundary conditions. 
The ground state energy of the model with periodic boundary conditions
is, then, given by
\ba
&&
E_{\rm periodic}= -L\int_{-\pi}^{\pi} dk \rho_{0} (k) \cos k.
\ea

The $O(1/L)$ contribution  $\rho_{1} (k)$ 
can be divided into three parts which reflect
the geometry (open boundary conditions),
the effect of the boundary fields, and
the existence of the boundary strings;
\ba
\rho_{1} (k) & = & \rho_{\rm geom} (k) + \rho_{\rm field} (k)
+\rho_{\rm string}(k). 
\ea
We further separate $\rho_{\rm field}(k)$ to
$\rho_{\rm field} (k)=\rho_{\rm field}(k,h_{1})+\rho_{\rm field}(k,h_{L})$
where $\rho_{\rm field}(k,h_{l})$ is the 
contribution from the boundary field $h_{l}$,
and $\rho_{\rm str}(k)$ to
$\rho_{\rm str} (k)=\rho_{\rm str}(k,h_{1})+\rho_{\rm str}(k,h_{L})$
where $\rho_{\rm str}(k,h_{l})$ is the 
contribution from the boundary string corresponding to the 
boundary field $h_{l}$.

Here $\rho_{\rm geom}(k) $ is given by
\ba
&&
\rho_{\rm geom} (k) = -\frac{\cos k}{2\pi} \int_{-\infty}^{\infty} dw
    \frac{e^{-iw \sin k -c |w|/2}}{1+e^{c|w|}} 
-2 \cos k K(2\sin k),
\ea
and $\rho_{\rm field} (k,h)$ is given by
\ba
&&
\rho_{\rm field}(k,h) =
\frac{\xi (k,h)}{2\pi} -\frac{\cos k}{2\pi^{2}} \int_{-\pi}^{\pi} dk'
\int_{-\infty}^{\infty} dw \frac{e^{-2iw(\sin k -\sin k')-c|w|}}{\cosh (cw)}.
\ea 
The form of $\rho_{\rm str}(k,h)$ is depend on 
the value of the boundary field $h$.

The energy in the case that there are boundary strings, is given by
\ba
&&
E = -L\int_{-\pi}^{\pi} dk \rho_{0} (k) \cos k
            -\int_{-\pi}^{\pi} dk \rho_1 (k)\cos k
           -2(\cosh \chi_1 +\cosh \chi_L ).
\ea

Then the energy difference between 
the configuration which contains boundary string 
and the configuration which does not contain the boundary string,
corresponding to the boundary field $h_l$, is given by 
\ba 
&&
\Delta E(h_{l}) = -2 \cosh \chi_{l} + e(h_{l}),
\ea
where
\ba
&&
e(h_{l}) = -\int_{-\pi}^{\pi} dk \rho_{\rm str}(k,h_{l})\cos k.
\ea

If the strength of the boundary field $h_{l}$ is 
such that $c<|2 \sinh \chi_{l}|$,
then 
\ba
\rho_{\rm str} (k,h_l) 
& = & 4c \cos k\int_{-\infty}^{\infty} d\Lam
\frac{ K \left(\frac{2c\Lam}{2 \sinh \chi_{l} -c} \right) 
K (2\sin k -2\Lam) }{
2 \sinh \chi_{l} -c },
\ea
and 
\ba
e(h_{l}) & =& - \oL  \int_{-\pi}^{\pi} dk \int_{-\infty}^{\infty} d\Lam
\frac{4c \cos^{2} k}{2\sinh \chi_{l} -c} 
K (\frac{2c\Lam}{2\sinh \chi_{l} -c})
K (2\sin k -2\Lam).
\ea
Noticing $K(x)>0$ ($x$ is real), we see that $e(h_{l})<0$.

On the other hand, if  $|2\sinh \chi_{l}| <c$, then 
\ba
\rho_{\rm str} (k,h_l) & = & 
\frac{\cos k}{4\pi} \left[ G(\frac{c}{2} -\sinh \chi_{l},k)
+ G(\frac{c}{2} +\sinh \chi_{l} ,k) \right],
\ea
where
\ba 
G(x,y) & =&  \int_{-\infty}^{\infty} dw
\frac{e^{-iw \sin y-x|w|}}{\cosh cw/2} 
= \frac{x}{2\pi} \int_{-\infty}^{\infty} d\Lam 
\frac{K(c(\Lam-\sin y)/x)}{\cosh (\pi \Lam /c)}.
\ea
In this case, $e(h_{l})$ is given by
\ba
e(h_{l}) & =& - \oL \int_{-\pi}^{\pi} dk 
\frac{\cos^{2} k}{4\pi} \left[ G(\frac{c}{2} -\sinh \chi_{l},k)
+ G(\frac{c}{2} +\sinh \chi_{l},k) \right].
\ea
Again, noticing $G(x,y)>0$ ($x>0$), we see that $e(h_{l})<0$.

We thus conclude that the configuration which
contains boundary strings has lower energy
than the configuration which does not contain the
boundary strings. 
Therefore, if $h_{l} >1$ for $l=1,L$, 
the ground state rapidity configuration of the open 
Hubbard chain with boundary fields at the half-filling 
contains the boundary strings for 
each boundary fields. 

The surface energy of the Hubbard chain with boundary fields
\ba
&&
E_{\rm surface} = E^{\rm gr} -E^{\rm gr}_{\rm periodic},
\ea
where $E^{\rm gr}$ is the ground state energy of the 
Hamiltonian $H$ and $E^{\rm gr}_{\rm periodic}$ 
is the ground state energy of the model with periodic boundary 
conditions,
is given by
\ba
E_{\rm surface} & = & -\int_{-\infty}^{\infty} dk 
[\rho_{\rm geom}(k) + \rho_{\rm field}(k) +\rho_{\rm str}(k)]\cos k \nn \\
&&
-2 (\cosh \chi_1 +\cosh \chi_L).
\ea

\noindent
{\it Conclusion}

Although, we did not yet performed to classify 
the all solutions of the Bethe ansatz equations 
\eq{bethe-eq1} and \eq{bethe-eq2},
we have found that there exist the boundary string solutions in addition to 
the bulk strings.

It was known that the open Hubbard chain
with another type of the boundary fields 
$
H_{\mbox{{\scriptsize boundary}}}'
=
-h_{1}(n_{1\uparrow}-n_{1\downarrow})
-h_{L}(n_{L\uparrow}-n_{L\downarrow})
$
can be solved.
The related Bethe ansatz equations for this 
Hamiltonian 
contain factors which depend both the coupling constant
and boundary fields \cite{SW,DY}.
The analysis of these Bethe ansatz equations 
will be appeared elsewhere.

After this letter was written up, we find 
the preprint which relates to our results \cite{BF}.
In their paper, boundary strings for the both
charge and spin rapidities are studied.
 
 
\vspace{24pt}


\end{document}